\documentstyle[12pt,a4,graphicx]{article}

\newcommand{\be}{\begin{equation}}
\newcommand{\ee}{\end{equation}}
\newcommand{\ba}{\begin{eqnarray}}
\newcommand{\ea}{\end{eqnarray}}

\newcommand{\re}{\mbox{Re}\,}
\newcommand{\im}{\mbox{Im}\,}

\begin{document}
\begin{titlepage}
\begin{flushright}
CAFPE-64/05\\
IFIC/05-49\\
UG-FT-194/05
\end{flushright}
\vspace{2cm}
\begin{center}

{\large\bf Charged Kaon $K^+\to3\pi$ CP Violating 
Asymmetries vs $\varepsilon_K'/\varepsilon_K$ 
\footnote{Work supported in part  
by MEC, Spain and FEDER, European Commission (EC)
 Grant Nos. FPA2003-09298-C02-01 (J.P.)
and FPA2004-00996 (I.S.), 
by Junta de Andaluc\'{\i}a Grant No. FQM-101 (J.P.),
and by the EC RTN EURIDICE Contract No. HPRN-CT-2002-00311 (J.P. and I.S.).
E.G. is indebted to the EC for the Marie Curie
Fellowship No. MEIF-CT-2003-501309.
Invited talk given by J.P.
at ``XII International Conference on Quantum Chromodynamics
 (QCD '05)'', 4-8 July 2005, Montpellier, France.}}\\
\vfill
{\bf Joaquim Prades$^{a)}$, Elvira G\'amiz$^{b)}$
 and Ignazio Scimemi$^{c)}$}\\[0.5cm]
$^{a)}$ Centro Andaluz de F\'{\i}sica de las Part\'{\i}culas
Elementales (CAFPE) and Departamento de
 F\'{\i}sica Te\'orica y del Cosmos, Universidad de Granada \\
Campus de Fuente Nueva, E-18002 Granada, Spain.\\[0.5cm]

$^{b)}$ Department of Physics and Astronomy, The University of
Glasgow, \\  Glasgow G12 8QQ, United Kingdom.\\[0.5cm]

$^{c)}$  Departament de F\'{\i}sica
Te\`orica, IFIC, Universitat de Val\`encia-CSIC,\\
Apt. de Correus 22085, E-46071 Val\`encia, Spain.\\[0.5cm]

\end{center}
\vfill
\begin{abstract}
\noindent
We present the 
next-to-leading order full results in Chiral Perturbation Theory
for the charged Kaon $K\to3\pi$ slope $g$ CP violating asymmetries.
We discuss the constraints that a measurement of these asymmetries
would impose on the Standard Model results of $\varepsilon_K'$
and search for new physics. We also study the kind of information
that
such measurement can provide on Im $G_8$, Im $(e^2 G_E)$ and higher order
weak couplings.
\end{abstract}
\vfill
September 2005
\end{titlepage}

\section{Introduction and Motivation}
Direct CP violation in Kaons has been unambiguously
established in $K\to\pi\pi$ decays  by KTeV \cite{KTeV}
 and NA48 \cite{NA48} through the measurement of
Re $(\varepsilon_K'/\varepsilon_K)$, which present world
average is \cite{KTeV,NA48,NA31,E731}
\be
\label{epsprime}
{\rm Re} \, \left(\frac{\varepsilon_K'}
{\varepsilon_K}\right) = (1.67 \pm 0.16) \cdot 10^{-3} \, . 
\ee
The theoretical understanding of this quantity within the
Standard Model (SM) is not at the same level.
We just mention here the most recent advances:
the Chiral Perturbation Theory (CHPT) calculation
\cite{KMW90,K2piNLO} and the isospin breaking corrections 
\cite{CENP03}
are both fully known to next-to-leading order (NLO)
and the r\^ole of final state interactions (FSI) has also been
understood \cite{FSI} --for a more extensive description of these 
works and references see \cite{Toni}.
There also have been recent advances on the calculation of the 
lowest order CHPT couplings Im $G_8$ and Im $(e^2 G_E)$
\cite{elmatrix,strmatrix,matrix,latmatrix}. They are not fully
under control yet  and more work is needed.

Asymmetries in the Dalitz variable slope $g$ of $K\to3\pi$
offer a very promising opportunity to study direct CP
violation in Kaon decays. In fact, there are several experiments,
NA48/2 \cite{KEKE04} at CERN, KLOE \cite{KLOE} at Frascati
and OKA \cite{OKA} at Protvino which have announced an expected
sensitivity to these asymmetries of the order of $10^{-4}$, 
i.e., one order of magnitude better than at present \cite{AJI03}.
On the theory side, though the first NLO in CHPT calculation
of $K\to3\pi$ was done long ago \cite{KMW90}, the analytical results
were unfortunately  not available until recently \cite{BDP03,GPS03}.
$K\to3\pi$ CP violating asymmetries were therefore predicted just
at LO in CHPT plus various estimates of NLO effects \cite{previous}.
The first full NLO calculation within CHPT for those asymmetries
was done in \cite{GPS03}.

\section{Technique}

The effective  quantum field theory of the SM  at energies below
or of the order of 1 GeV is CHPT \cite{CHPT}. Some introductory
lectures on CHPT can be found in \cite{lectures} while some
recent reviews in \cite{reviews}.

Recently, the  one-loop in CHPT $K\to3\pi$ calculation 
was redone by two groups \cite{BDP03,GPS03} in the isospin limit
as mentioned above. All the needed notation and definitions were
given there. There are also now available the NLO
$K\to3\pi$ isospin breaking effects from quark masses and electromagnetic
interactions \cite{isoK3pi}. These effects turn out to be very suppressed
and certainly much smaller than the contribution of the  unknown
counterterms, we therefore use the isospin limit results.
Notice that some misprints in the first reference in \cite{GPS03}
were reported in the third reference in \cite{GPS03}.

 There appear eleven unknown counterterms at NLO in the isospin limit.
The real part of them and the LO couplings $G_8$ and $G_{27}$
can be fixed from a fit to all available $K\to\pi\pi$ amplitudes
at NLO in CHPT \cite{K2piNLO} 
and $K\to3\pi$ amplitudes and slopes also at NLO \cite{BDP03,GPS03}.
This was done in \cite{BDP03} and we used the results of the fit 
 as inputs in all the results we report here.

 The values we use for Im $(e^2 G_E)$ and Im $G_8$ can be found
in \cite{GPS03}. They are taken mainly from \cite{elmatrix,strmatrix}
 but are compatible also with those in \cite{matrix,latmatrix}.

 The imaginary part of the order $p^4$ counterterms, 
Im $\widetilde K_i$, are much more problematic to predict. They cannot
be obtained from data and there is not a NLO in $1/N_c$
($N_c$ is the number of QCD colors)
available calculation   for them. One can still get the order of 
magnitude and/or signs of Im $\widetilde K_i$ using several approaches.
We followed in \cite{GPS03} a more naive approach but that is enough
for the purpose of estimating the effect of those counterterms.
Namely, we assumed that the ratio of the real to the imaginary part
of the octet couplings  is roughly  dominated 
by the same strong dynamics  at LO and at NLO in CHPT, i.e.
\ba
\frac{\im \widetilde K_i}{\re \widetilde K_i}
&\simeq& \frac{\im G_8}{\re G_8} 
\simeq \frac{\im G_8'}{ \re G_8'} \simeq (0.9 \pm 0.3) \, \im \tau
\nonumber \\ 
&=& - (0.9\pm0.3) \,\,
\im \left(\frac{V_{td}V_{ts}^*}{V_{ud}V_{us}^*}\right).
\ea

\section{$K\to3\pi$ CP Violating Asymmetries}

The definition of the CP violating  asymmetries
in the slope $g$ and analogous asymmetries for the decay rates
$\Gamma$ can be found, for instance, in \cite{GPS03}. They
start at ${\cal O}(p^2)$ in CHPT and at NLO require final state
interactions  phases of the three-pions at NLO too, i.e.,
an ${\cal O}(p^6)$ calculation.

Though the  full ${\cal O}(p^6)$ result is not yet
available, we calculated analytically the 
full two-pion cut diagrams contribution to the FSI phases
for the charged Kaon decays in \cite{GPS03}. In \cite{GPS05},
we present the full analytic results 
for the  neutral Kaon decays 
including the three-pion cuts  both  for the neutral and
charged Kaon decays.
 We also discuss there some applications of this $K\to3\pi$ 
FSI calculation to the Cabibbo's proposal to measure the
$a_0-a_2$ pion scattering lengths combination \cite{CAB04}
 from the cusp effect in the $\pi^0\pi^0$ spectrum
in $K^+\to\pi^+\pi^0\pi^0$ and $K_L\to \pi^0\pi^0\pi^0$ decays.

Including the calculated NLO FSI and substituting the pion and 
Kaon masses,  Re $G_8$, $G_{27}$ and the real part of the NLO
CHPT couplings, the result we get for the asymmetry $\Delta g_C$
of the slope $g_C$ in $K^+\to\pi^+\pi^+\pi^-$ decay
is
\ba
\label{deltagc}
\frac{\Delta g_C}{10^{-2}} &\simeq&
\left[ \left(0.7\pm0.1\right) {\rm Im} G_8 + \left(4.3\pm 1.6\right) 
 {\rm Im} \widetilde K_2 \right.  \\
 &-& \left.(18.1 \pm 2.2) {\rm Im} \widetilde K_3
-(0.07 \pm0.02) {\rm Im} (e^2 G_E) \right] \, . \nonumber
\ea

And, when values for  the imaginary part of the  needed couplings
are taken as explained in the previous section, one gets
\be
\Delta g_C=-(2.4\pm1.2) \cdot 10^{-5} \, .
\ee
Results for the rest of the asymmetries can be found in \cite{GPS03}.

\section{$K\to3\pi$ CP Violating Asymmetries vs $\varepsilon_K'$}

\begin{figure}
\begin{center}
\begin{minipage}[t]{1cm}\vskip-4.cm 
$\frac{{\rm Im} G_8}{{\rm Im} \tau}$\end{minipage}
\hspace*{-.5cm}\includegraphics[height=0.45\textwidth]{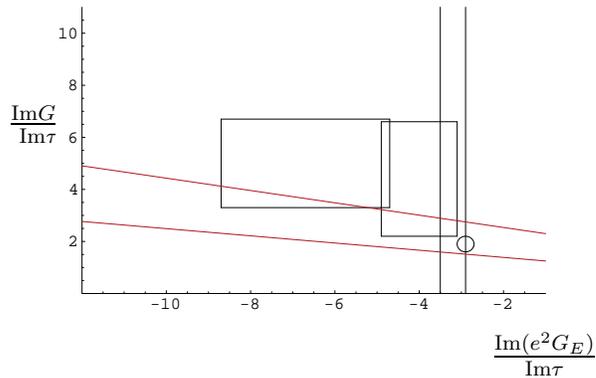}
\begin{minipage}[b]{1cm}\hspace*{-1cm}\vspace{.4cm}
$\frac{{\rm Im} (e^2 G_E)}{{\rm Im} \tau}$\end{minipage}
\caption{$\varepsilon_K'$: Theory vs Experiment. See text for explanation.}
\label{fig:status}
\end{center}
\end{figure}

Including FSI to all orders  and  CHPT and isospin breaking at NLO
\cite{K2piNLO,CENP03,FSI} one gets within the Standard Model,
\be
\label{theps}
{\rm Re} \, \left(
\frac{\varepsilon_K'}{\varepsilon_K} \right) \simeq
-\left[ \left( 1.88\pm1.0\right) {\rm Im} G_8 +  \left(0.38\pm0.13
\right) {\rm Im} (e^2 G_E) \right] \, .
\ee
Using this result, the experimental measurement in (\ref{epsprime})
imposes that Im $G_8$ and Im $(e^2 G_E)$ are constrained to be within
the horizontal band in Figure \ref{fig:status}.
In the same figure we also show the present theoretical predictions 
at NLO in $1/N_c$ 
expansion for   Im $G_8$ and Im $(e^2 G_E)$:  
from \cite{elmatrix,strmatrix} 
--rectangle on the right, from \cite{matrix} --rectangle on the left,
and  from \cite{latmatrix} --vertical band.
The circle is the leading in $1/N_c$ prediction.

\begin{figure}
\begin{center}
\begin{minipage}[t]{1cm}\vskip-4.cm 
$\frac{{\rm Im} G_8}{{\rm Im} \tau}$\end{minipage}
\hspace*{-.5cm}\includegraphics[height=0.45\textwidth]{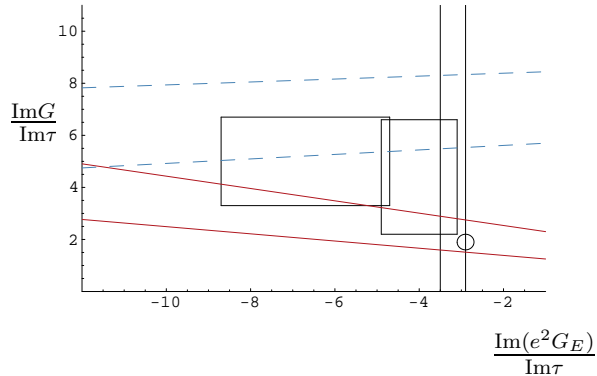}
\begin{minipage}[b]{1cm}\hspace*{-1cm}\vspace{.4cm}
$\frac{{\rm Im} (e^2 G_E)}{{\rm Im} \tau}$\end{minipage}
\caption{$\varepsilon_K'$ vs $\Delta g_C$ for  $\Delta g_C =
- 3.5 \cdot 10^{-5}$.}
\label{fig:gclim}
\end{center}
\end{figure}

\begin{figure}
\begin{center}
\begin{minipage}[t]{1cm}\vskip-4.cm 
$\frac{{\rm Im} G_8}{{\rm Im} \tau}$\end{minipage}
\hspace*{-.5cm}\includegraphics[height=0.45\textwidth]{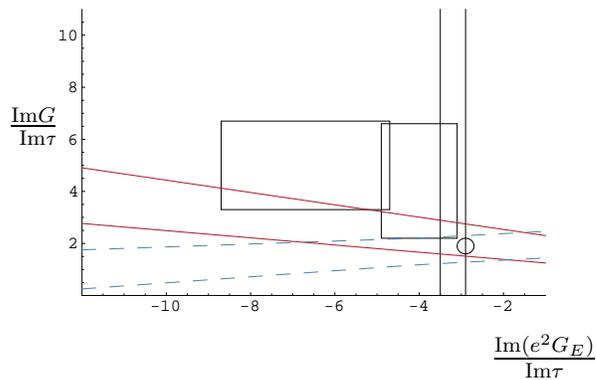}
\begin{minipage}[b]{1cm}\hspace*{-1cm}\vspace{.4cm}
$\frac{{\rm Im} (e^2 G_E)}{{\rm Im} \tau}$\end{minipage}
\caption{$\varepsilon_K'$ vs $\Delta g_C$ for  $\Delta g_C =
- 1 \cdot 10^{-5}$.}
\label{fig:figc}
\end{center}
\end{figure}

A measurement of $\Delta g_C$ can have an important impact on 
constraining what  we know on Im $G_8$ and Im $(e^2 G_E)$ from
$\varepsilon_K'$. To assess the quality of these constraints, we plot
in Figure \ref{fig:gclim}  the dashed horizontal band  that one gets
using (\ref{deltagc})  for $\Delta g_C = -3.5 \cdot 10^{-5}$ 
with a typical 
(20 $\sim$ 30)\% uncertainty together with  the $\varepsilon_K'$ 
constraints shown in Figure \ref{fig:status}.
In Figure \ref{fig:figc},  we show the same information as in
Figure \ref{fig:gclim} but for  $\Delta g_C = -1 \cdot 10^{-5}$
with the same (20 $\sim$ 30)\% typical uncertainty.

\section{Conclusions}

The CP violating asymmetry $\Delta g_C$ is dominated
by the value of Im $G_8$ and its final uncertainty comes mainly
from this input. This is the only $K\to3\pi$ CP asymmetry with 
uncertainty below 50\%. The predictions for the rest of the asymmetries
studied can be found in \cite{GPS03}.

The eventual measurement of the  asymmetry
$\Delta g_C$ when combined with the accurate measurement of 
$\varepsilon_K'$ \cite{KTeV,NA48,NA31,E731}  in (\ref{epsprime})
offers an opportunity to both check the Standard Model calculations
 and to obtain more information on possible new physics.

The SM prefers values for $\Delta g_C$ larger than $-0.4 \cdot 10^{-4}$
and an experimental bound of the order or smaller than
$-2 \cdot 10^{-4}$ would indicate the presence of new physics 
-- see Figures \ref{fig:gclim} and \ref{fig:figc}.
For a discussion on possible SUSY implications of a measurement
of these asymmetries see \cite{AIM00}.

The  CP asymmetries $\Delta g_N$ and in the decay rate were also
discussed in \cite{GPS03} and we found that they are dominated  by the
imaginary part of the ${\cal O} (p^4)$ counterterms. Information
on  these asymmetries would therefore give very interesting
constraints  about the size of those imaginary parts.

As a final remark, direct CP violating asymmetries in $K\to3\pi$
provide  extremely interesting and valuable information on the SM and 
its possible extensions which is complementary to the one 
obtained from $\varepsilon_K'$. The first experimental result
by NA48/2 at CERN was already announced \cite{NA482}
\be
|\Delta g_C| = (0.5\pm3.8) \cdot 10^{-4}, \, 
\ee
which uncertainty is expected to be reduced.

\vspace*{-0.3cm}
\section*{Acknowledgments}
It is a pleasure to thank Stephan Narison for the invitation
to this very enjoyable conference. 
 J.P. also thanks the warm hospitality
 of  the Department of Theoretical Physics at Lund University where
 this work was written.

\end{document}